\documentclass[epsf,useAMS,usenatbib,usegraphicx]{mn2e}
\usepackage{epsfig}
\usepackage{natbib}
\usepackage{aas_macros} 
\usepackage{amsmath}
\usepackage{amssymb}
\usepackage{appendix}
\usepackage{booktabs}
\usepackage{upgreek}
\usepackage{graphicx}
\usepackage{url}

\title[KIC\,4768731]{KIC\,4768731: a bright long-period roAp star in the {\it Kepler} Field}

\author[Barry Smalley et al.] 
{B. Smalley$^{1,}$\thanks{Email: b.smalley@keele.ac.uk}, 
E. Niemczura$^{2}$,
S.~J. Murphy$^{3,4}$,
H. Lehmann$^{5}$,
D.~W. Kurtz$^{6}$, \newauthor
D.~L. Holdsworth$^{1}$,
M.~S. Cunha$^{7}$,
L.~A. Balona$^{8}$,
M. Briquet$^{9}$,
H. Bruntt$^{10}$,\newauthor
P. De Cat$^{11}$,
P. Lampens$^{11}$,
A.~O. Thygesen$^{12}$,
K. Uytterhoeven$^{13,14}$\\
$^{1}$Astrophysics Group, Lennard-Jones Laboratories, Keele University,
Staffordshire ST5 5BG, United Kingdom\\
$^{2}$Instytut Astronomiczny, Uniwersytet Wroc{\l}awski, Kopernika 11, 51-622
Wroc{\l}aw, Poland\\
$^{3}$Sydney Institute for Astronomy (SIfA), School of Physics, University of Sydney
NSW 2006, Australia\\
$^{4}$Stellar Astrophysics Centre, Department of Physics and Astronomy, Aarhus University, 8000 Aarhus C, Denmark\\
$^{5}$Th\"{u}ringer Landessternwarte Tautenburg (TLS), Sternwarte 5, 07778
Tautenburg, Germany\\
$^{6}$Jeremiah Horrocks Institute, University of Central Lancashire, Preston PR1
2HE, United Kingdom\\
$^{7}$Instituto de Astrofisica e Ci\^{e}ncias do Espa\c{c}o, Universidade do Porto,
CAUP, Rua das Estrelas, PT4150-762, Porto, Portugal\\
$^{8}$South African Astronomical Observatory, P.O. Box 9, Observatory 7935, Cape Town, South Africa\\
$^{9}$Institut d'Astrophysique et de G\'{e}ophysique, Universit\'{e} de Li\`{e}ge,
All\'{e}e du 6 Ao\^{u}t 19C, B-4000 Li\`{e}ge, Belgium\\
$^{10}$Stellar Astrophysics Center, Department of Physics and Astronomy, Aarhus
University, Ny Munkegade 120, DK-8000 Aarhus C, Denmark\\
$^{11}$Royal observatory of Belgium, Ringlaan 3, B-1180 Brussel, Belgium\\
$^{12}$18 Zentrum f\"{u}r Astronomie der Universit\"{a}t Heidelberg,
Landessternwarte, K\"{o}nigstuhl 12, 69117 Heidelberg, Germany\\
$^{13}$Instituto de Astrofisica de Canarias, E-38205 La Laguna, Tenerife,
Spain\\
$^{14}$Universidad de La Laguna, Departamento de Astrofisica, E-38206 La Laguna, Tenerife, Spain
}

\date{Accepted 2015 July 6.  Received 2015 July 3; in original form 2015 June 4} 

\begin{document}

\maketitle 

\begin{abstract}

We report the identification of 61.45~d$^{-1}$ (711.2~$\upmu$Hz)
oscillations, with amplitudes of 62.6-$\upmu$mag, in KIC\,4768731 (HD\,225914)
using {\it Kepler} photometry. This relatively bright ($V$=9.17) chemically
peculiar star with spectral type A5\,Vp~SrCr(Eu) has previously been found to
exhibit rotational modulation with a period of 5.21~d. Fourier analysis
reveals a simple dipole pulsator with an amplitude that has remained stable over
a 4-yr time span, but with a frequency that is variable. Analysis of
high-resolution spectra yields stellar parameters of $T_{\rm eff} = 8100 \pm
200$~K, $\log g = 4.0 \pm 0.2$, [Fe/H] = $+0.31 \pm 0.24$ and $v \sin i = 14.8
\pm 1.6$~km\,s$^{-1}$. Line profile variations caused by rotation are also evident.
Lines of Sr, Cr, Eu, Mg and Si are strongest when the star is brightest, while Y
and Ba vary in anti-phase with the other elements. The abundances of rare
earth elements are only modestly enhanced compared to other roAp stars of
similar $T_{\rm eff}$ and $\log g$. Radial velocities in the literature suggest
a significant change over the past 30\,yr, but the radial velocities presented
here show no significant change over a period of 4\,yr.

\end{abstract}


\section{Introduction}

The chemically peculiar Ap stars are a spectroscopic subclass of A-type stars
which exhibit strong enhancements of one or more elements, notably Cr, Eu, Si
and Sr \citep{1933ApJ....77..330M}. Strong magnetic fields in Ap stars were
found by \cite{1947ApJ...105..105B} and their presence appears ubiquitous with
field strengths reaching several kilogauss \citep{2009ARA&A..47..333D}. As a
group the Ap stars rotate more slowly than normal A-type stars due to magnetic
breaking \citep{2000A&A...353..227S} and exhibit photometric and spectral
variability due to inhomogeneous surface distributions of elemental abundances
concentrated into spots by the magnetic field
\citep{1983aspp.book.....W,1996Ap&SS.237...77S}.

\begin{table*}
\caption{Catalogue of roAp stars.  The first two columns are the HD number
and other name if available.  The effective temperature (K) is obtained from
the Balmer line profiles when available.  The luminosity, $L/$L$_{\sun}$ is
obtained from $T_{\rm eff}$ and the surface gravity when available or from
the radius.  The projected rotational velocity, $v\sin i$ (km\,s$^{-1}$) is
given.  The averaged magnetic field, $\bar{B}$ (kG), from \citet{2009MNRAS.394.1338B}
are shown; values in italics are other measures of $B$.  The rotation
period, $P_{\rm rot}$ is determined from the light curve.  A maximum of four
pulsation frequencies is given.}
\label{roApstars}
\scriptsize
\begin{tabular}{rlrrrrrll}
\hline
\multicolumn{1}{c}{HD} &
\multicolumn{1}{l}{Name} &
\multicolumn{1}{c}{$\log T_{\rm eff}$} &
\multicolumn{1}{c}{$\log L$} &
\multicolumn{1}{c}{$v\sin i$} &
\multicolumn{1}{c}{$\bar{B}$} &
\multicolumn{1}{c}{$P_{\rm rot}$} &
\multicolumn{1}{c}{Frequencies} &
\multicolumn{1}{l}{Reference} \\
\multicolumn{1}{c}{} &
\multicolumn{1}{l}{} &
\multicolumn{1}{c}{K} &
\multicolumn{1}{c}{L$_{\sun}$} &
\multicolumn{1}{c}{km\,s$^{-1}$} &
\multicolumn{1}{c}{kG} &
\multicolumn{1}{c}{d} &
\multicolumn{1}{c}{$\upmu$Hz} &
\multicolumn{1}{l}{} \\
\hline
   6532 & AP~Scl       & 3.914 & 1.22 &   30.0  & 0.40 &  1.944973 &  2390.2, 2396.2, 2402.2, 2408.1, &  \citet{1996MNRAS.281..883K} \\
   9289 & BW~Cet       & 3.907 & 1.08 &   10.5  & 0.07 &  8.55     &  1585.1, 1554.8, 1605.4          &  \citet{1994MNRAS.271..421K} \\
  12098 & V988~Cas     & 3.892 &{0.88}&         & 1.00 &  5.460    &  2173.7, 2164.2, 2180.6, 2305.6  &  {\citet{2001A&A...380..142G}} \\
  12932 & BN~Cet       & 3.884 &{1.21}&    2.5  & 0.64 &  3.5295?  &  1436.3                          &  \citet{1994MNRAS.271..305M} \\
  19918 & BT~Hyi       & 3.899 & 1.06 &    3.0  & 0.21 &           &  1510.2, 3020.1, 1480.7          &  \citet{1995MNRAS.276.1435M} \\
  24355 & J0353        & 3.916 &      &         &      & 13.859611 &  2596                            &  \citet{2014MNRAS.439.2078H} \\
  24712 & HR~1217      & 3.860 & 0.89 &         & 0.76 &  12.45877 &  2619.5, 2653.0, 2687.6, 2721.0, &  \citet{2002MNRAS.330L..57K} \\
  42659 & UV~Lep       & 3.900 & 1.48 &   19.0  & 0.39 &           &  1735.5                          &  \citet{1993IBVS.3844....1M} \\
  60435 & V409~Car     & 3.910 & 1.14 &   10.8  & 0.30 &  7.6793   &   709.0,  761.4,  842.8,  939.7, &  \citet{1987ApJ...313..782M} \\
  69013 &              & 3.881 & 0.60 &    4.0  & {\it 4.8} &           &  1485                            &  \citet{2013MNRAS.431.2808K}  \\
  75445 &              & 3.886 & 1.17 &    2.0  &  -   &           &  1850.0                          &  {\citet{2009A&A...493L..45K}}  \\
  80316 & LX~Hya       & 3.918 & 1.05 &   32.0  &  -   &  2.08860  &  2246.1, 2251.6, 2257.2          &  \citet{1997MNRAS.289..645K}      \\
  83368 & HR~3831      & 3.877 & 1.09 &   33.0  & 0.81 &  2.851982 &  1415.8, 1419.9, 1424.0, 1428.0, &  \citet{1997MNRAS.287...69K}     \\
  84041 & AI~Ant       & 3.916 &      &   25.0  & 0.48 &  3.69     &  1113.0, 1085.0, 1145.0          &  \citet{1993MNRAS.263..273M}  \\
  86181 & V437~Car     & 3.865 & 1.05 &         & 0.40 &           &  2688.0                          &  \citet{1994IBVS.4013....1K}     \\
  92499 &              & 3.875 & 1.14 &    3.3  & {\it 8.2} &           &  1602                            &  \citet{2010MNRAS.404L.104E}      \\
  96237 & TX~Crt       & 3.892 & 0.88 &    6.0  & {\it 2.9} &           &  1200                            &  \citet{2013MNRAS.431.2808K}  \\
  99563 & XY~Crt       & 3.886 & 1.10 &   28.0  & 0.57 &           &  1553.7, 1561.6, 1545.7, 1557.6, &  \citet{2006MNRAS.366..257H}   \\
 101065 & V816~Cen     & 3.810 & 0.91 &    4.0  & 1.02 &           &  1372.8, 1381.5, 1314.6, 1379.8, &  {\citet{2008A&A...490.1109M}} \\
 115226 &              & 3.883 & 0.86 &   27.0  & 0.74 &           &  1534.0                          &  {\citet{2008A&A...479L..29K}}  \\
 116114 &              & 3.870 & 1.32 &    2.2  & 1.92 &           &   790.0                          &  \citet{2005MNRAS.358..665E}      \\
 119027 & LZ~Hya       & 3.875 & 0.79 &         &      &           &  1953.7, 1940.5, 1913.4, 1887.9, &  \citet{1998MNRAS.300..188M}   \\
 122970 & PP~Vir       & 3.840 & 0.82 &    4.2  & 0.19 &   3.877   &  1502.5, 1477.8, 1476.8, 1478.9  &  \citet{2002MNRAS.330..153H}   \\
 128898 & $\alpha$~Cir & 3.875 & 1.03 &   13.5  & 0.32 &   4.4792  &  2442.6, 2265.4, 2341.8, 2366.5, &  \citet{1994MNRAS.270..674K}     \\
 132205 &              & 3.892 & 0.77 &    9.5  & {\it 5.2}  &           &  2334                            &  \citet{2013MNRAS.431.2808K}  \\
 134214 & HI~Lib       & 3.858 & 0.85 &    2.6  & 0.46 &   248.00  &  2947.0, 2784.0, 2644.0, 2842.0, &  \citet{2007MNRAS.381.1301K}     \\
 137909 & $\beta$~CrB  & 3.908 & 1.37 &    3.5  & 0.51 &           &   967.0, 1062.0                  &  \citet{2008CoSka..38..423K} \\
 137949 &  33~Lib      & 3.869 & 1.09 &    3.0  & 2.14 &           &  2014.8, 4029.6, 1769.0          &  \citet{2005MNRAS.358L...6K}     \\ 
 143487 &              & 3.845 & 0.0  &    1.5  & {\it 4.7} &           &  1730                            &  \citet{2013MNRAS.431.2808K}  \\
 148593 &              & 3.894 & 0.78 &    5.0  & {\it 3.0} &           &  1560                            &  \citet{2013MNRAS.431.2808K}  \\
 150562 & V835~Ara     & 3.9:  &      &    1.5  & {\it 5.0} &           &  1550.0                          &  \citet{1992IBVS.3750....1M}   \\
 151860 &              & 3.848 & 0.42 &    4.5  & {\it 2.5} &           &  1355                            &  \citet{2013MNRAS.431.2808K}  \\
 154708 &              & 3.829 & 0.73 &    4.0  & 6.54 &   5.3666  &  2088.0                          &  \citet{2006MNRAS.372..286K}     \\ 
 161459 &{V834~Ara}    &       &      &         & 1.76 &           &  1390.9                          &  \citet{1990IBVS.3507....1M}  \\
 166473 & V694~CrA     &{3.889}&{1.22}&    2.5  & 2.15 &           &  1833.0, 1886.0, 1928.0          &  \citet{2007MNRAS.380..181M} \\ 
 176232 & 10~Aql       & 3.899 & 1.32 &    2.7  & 0.46 &           &  1447.9, 1396.9, 1427.1, 1366.2, &  {\citet{2008A&A...483..239H}}      \\
 177765 &              & 3.903 & 1.54 &    2.5  & {\it 3.6} &           &   706                            &  \citet{2012MNRAS.421L..82A}   \\
 185256 & V4373~Sgr    &       &      &    6.2  & 0.71 &           &  1630.0                          &  \citet{1995IBVS.4209....1K}     \\
 190290 & CK~Oct       &       &      &   16.0  & 3.00 &  4.03     &  2270.0, 2230.0                  &  \citet{1991MNRAS.250..666M}   \\
 193756 & QR~Tel       &       &      &   17.0  & 0.36 &           &  1284.0                          &  \citet{1990IBVS.3509....1M}  \\
 196470 & AW~Cap       &       &      &         & 1.47 &           &  1544.0                          &  \citet{1990IBVS.3506....1M}  \\
 201601 & $\gamma$~Equ & 3.878 & 1.10 &    2.5  & 0.79 &  1785.70  &  1364.6, 1365.4, 1427.1, 1388.9, &  {\citet{2008A&A...480..223G}}\\ 
 203932 & BI~Mic       &       &      &    4.7  & 0.25 &           &  1280.5, 2838.0, 2772.3, 2737.3  &  \citet{1990MNRAS.246..699M}  \\
 213637 & MM~Aqr       &{3.822}&{0.64}&    3.5  & 0.74 &           &  1452.3, 1410.9                  &  {\citet{1998A&A...334..606M}}  \\ 
 217522 & BP~Gru       & 3.816 & 0.85 &    2.7  & 0.69 &           &  1215.3, 1199.9, 2017.4          &  \citet{1991MNRAS.250..477K}     \\
 218495 & CN~Tuc       &{3.888}&{1.10}&   16.0  & 0.77 &           &  2240.0                          &  \citet{1990IBVS.3509....1M}  \\ 
 218994 &              & 3.881 & 1.06 &    5.2  &      &           &  1170.0                          &  \citet{2008MNRAS.384.1140G}   \\
 225914 & KIC~4768731  & 3.888 & 1.33 &   14.8  & {\it 2.7} &  5.205    &  711.2, 713.5, 709.0             &  This Work \\ 
        & J0008        & 3.863 &      &         &      &           &  1739                            &  \citet{2014MNRAS.439.2078H} \\   
 258048 & J0629        & 3.820 &      &         &      &           &  1962                            &  \citet{2014MNRAS.439.2078H} \\
        & J0651        & 3.869 &      &         &      &           &  1532                            &  \citet{2014MNRAS.439.2078H} \\
        & J0855        & 3.892 &      &         &      &  3.09     &  2283                            &  \citet{2014MNRAS.439.2078H} \\
  97127 & J1110        & 3.799 &      &         &      &           &  1234                            &  \citet{2014MNRAS.439.2078H} \\
        & J1430        & 3.851 &      &         &      &           &  2726                            &  \citet{2014MNRAS.439.2078H} \\
        & J1640        & 3.869 &      &         &      &  3.67     &  1758                            &  \citet{2014MNRAS.439.2078H} \\
        & KIC~7582608  &{3.940}&{1.13}& $<$4    & {\it 3.1} & 20.20 & 2103                            &  \citet{2014MNRAS.443.2049H} \\
        & J1921        & 3.792 &      &         &      &           &  1490                            &  \citet{2015PhDT..........H} \\
        & J1940        & 3.839 &      &         &      &  9.58     &  2042                            &  \citet{2014MNRAS.439.2078H} \\
        & KIC~8677585  & 3.863 & 0.80 &    4.2  &      &           &  1659.8, 1621.8, 1504.3, 1676.0  &  \citet{2011MNRAS.410..517B} \\
        & KIC~10195926 & 3.869 & 1.61 &   21.0  &      &  5.68459  &  972.6, 976.7, 974.6, 970.6      &  \citet{2011MNRAS.414.2550K} \\
        & KIC~10483436 & 3.869 & 0.84 &         &      &  4.303    &  1353.0, 1347.6, 1358.4, 1508.9  &  \citet{2011MNRAS.413.2651B} \\
\hline
\end{tabular}
\end{table*}

The rapidly oscillating Ap (roAp) stars are a subset of the magnetic Ap stars
\citep{1982MNRAS.200..807K}. These exhibit short-timescale variations with
periods between 5 and 25 minutes and amplitudes up to 0.01\,mag. The roAp stars
are relatively rare with 61 known to date (Table~\ref{roApstars}), compared to
around 2000 known Ap stars \citep{2009A&A...498..961R}. Most have been found
using ground-based photometry of already known Ap stars. Some, which do not
exhibit photometric variations detectable from the ground, have been found by
spectroscopic studies looking for radial velocity variations
\citep[e.g.][]{2002MNRAS.337L...1K,2005MNRAS.358.1100E}. Eleven of the 61 known
roAp stars have now been found using the SuperWASP archive with $V$-band
amplitudes $>$0.5\,mmag \citep{2014MNRAS.439.2078H,2015PhDT..........H}.

The launch of the {\it Kepler} spacecraft enabled the search for photometric
oscillations with amplitudes well below those detectable from the ground. Seven
stars previously classified as Ap stars were observed by {\it Kepler} to search
for rapid oscillations, but only one was found to exhibit roAp pulsations
\citep{2011MNRAS.410..517B}. Two further previously unknown Ap stars were
identified from their pulsations and subsequent spectral analyses
\citep{2011MNRAS.413.2651B,2011MNRAS.414.2550K}. In the spectroscopic survey of
117 A and F stars observed by {\it Kepler}, \cite{2015MNRAS.450.2764N} identified one new Ap
star (KIC\,4768731), and re-classified two of the stars by
\cite{2011MNRAS.410..517B} as not Ap (KIC\,8750029 and KIC\,9147002). All the
{\it Kepler}-discovered roAp stars have oscillation amplitudes below 0.1\,mmag,
while one initially discovered by SuperWASP located in the {\it Kepler} field
has an amplitude of over 1\,mmag \citep{2014MNRAS.443.2049H}. The star discussed
here, KIC\,4768731, is brighter and has a lower pulsation frequency than the
other known roAp stars in the {\it Kepler} field.

\section{KIC\,4768731}

Prior to the {\it Kepler} mission, KIC\,4768731 (HD\,225914, BD\,+39 3919) was a
rather anonymous $V = 9.17$ star \citep{2000A&A...355L..27H} with a spectral
type A7 \citep{1925AnHar.100...17C}. The star is part of the double system WDS
19484+3952 with a 12th magnitude star (KIC\,4768748) at 12{\arcsec}
\citep{2001AJ....122.3466M}. There has been very little change in separation or
position angle over a period of 74\,yr, suggesting that the two stars might be a
common proper motion pair. Recent proper motion catalogues
\citep{2010AJ....139.2440R,2013AJ....145...44Z} are inconclusive in this regard,
with both stars having proper motions that agree to within the error bars.
Unfortunately, the {\it Hipparcos} parallax \citep{1997ESASP1200.....P} of $\pi
= 8.68 \pm 7.80$~mas is too uncertain to provide any meaningful distance
constraint. The distance of $\sim$350~pc estimated from the stellar parameters
(see Sect.~\ref{params}) would imply a physical separation of at least 4000~au.
However, the {\it Kepler} Input Catalog estimates $T_{\rm eff}$ = 4220~K and 
$\log g$ = 1.55 for the companion star, suggesting that this it a background
K-type giant star. In Sect.~\ref{observations} spectroscopic observations
confirm this.

\section{Spectral Classification}
\label{SpecClass}

\begin{figure}
\includegraphics[width=\columnwidth]{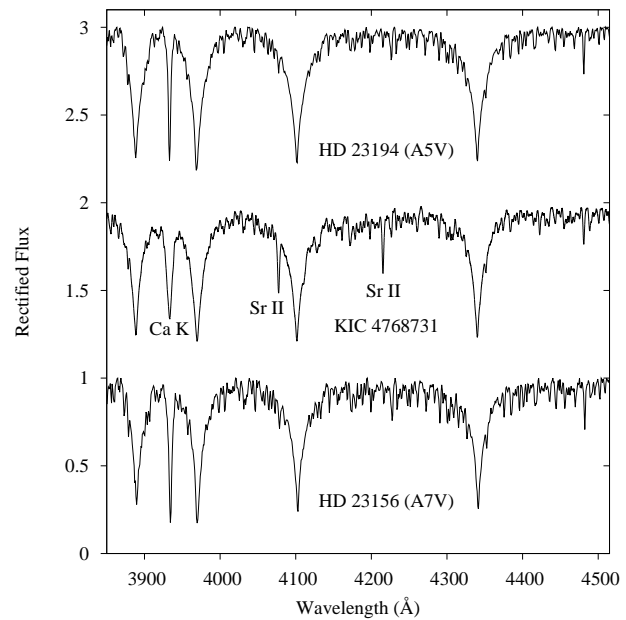}
\caption{Spectrum of KIC\,4768731 compared to the A5\,V and A7\,V MK standards.
Note the prominent Sr\,{\sc ii} lines at 4077{\AA} and 4216{\AA} and the
abnormal Ca\,{\sc ii} K line. The spectra have been offset for clarity.}
\label{fig:spectrum}
\end{figure}

KIC\,4768731 was included in the sample of several hundred A- and F-type stars 
to be observed at high-resolution as part of the {\it Kepler Asteroseismic
Science Consortium} (KASC) spectroscopic follow-up of stars with {\it Kepler}
light curves \citep{2015MNRAS.450.2764N}. A spectrum of this star was
taken on 2011 July 13 using the fibre-fed High Efficiency and Resolution
Mercator Echelle Spectrograph \citep[HERMES;][]{2011A&A...526A..69R}. Fig.\,\ref{fig:spectrum} presents the spectrum at
classification resolution. This spectrum is a smoothed version of the
high-resolution spectrum downgraded to a resolution of 1.8{\AA} to match that of
the spectral standards\footnote{Spectra of the MK standards were obtained at
R.O.~Gray's website:
\url{http://stellar.phys.appstate.edu/Standards/std1_8.html}}.

The H\,$\gamma$ line and the strength of the metal lines in general match that
of the A5V standard, but for an A5V star the Ca\,{\sc ii} K line is shallow and
abnormally broad. This abnormal shape is due to abundance stratification
within the star's atmosphere (see Sect.~\ref{lpv}). The Ca\,{\sc i} 4226{\AA} line, however, does not look
abnormal. Sr\,{\sc ii} lines are particularly strong (see the 4077 and
4216{\AA} lines). Cr lines are also strong: the Cr\,{\sc ii} 4111{\AA} line that
is blended in the redward core-wing boundary of H\,$\delta$ is pronounced, and
this is confirmed with the Cr\,{\sc ii} 3866{\AA} line. The Cr\,{\sc ii}
4172{\AA} line is mixed in with the Fe-Ti 4172--9{\AA} blend, but the 4172{\AA}
component is noticeably stronger. 

Other features to look for in the spectrum of Ap stars include strong Eu and Si
lines \citep[for an atlas of Ap star spectra, see][]{2009ssc..book.....G}. The
Eu\,{\sc ii} 4205{\AA} line is enhanced, but the Eu\,{\sc ii} 4130{\AA} line is
blended with the Si\,{\sc ii} doublet. That doublet of Si\,{\sc ii} (4128 and
4131{\AA}) is strong and, unlike in the comparison spectra, is more of a
blended feature than a separated doublet. This is not a product of rotation, as
KIC\,4768731 is clearly not a rapid rotator, but could reflect a strong Eu
line.  The Si\,{\sc ii} 3856{\AA} line appears strong, but other Si\,{\sc ii}
lines (4002, 4028 and 4076{\AA}) are not pronounced. These findings lead to the
classification A5\,Vp~SrCr(Eu).

\section{{\it Kepler} Observations}

\cite{2011A&A...529A..89D} classified KIC\,4768731 as exhibiting rotational
modulation with a frequency of 0.191206~d$^{-1}$ (5.23~d) and an amplitude of
5.407\,mmag. The AAVSO Variable Star Index \citep[VSX;][]{2006SASS...25...47W}
noted this star as ACV: (a suspected $\alpha^2$~CVn variable) with a period of
5.21~d and magnitude range of 10\,mmag based on {\it Kepler} light curves. The
VSX entry by R. Jansen was dated 2011 May based on a phase plot of {\it Kepler}
data from JD 2454964 to 2454998, noting also that the rise time from minimum to
maximum light is 50~per cent of the variable's period.

One month of Short Cadence (SC) data was acquired by KASC for this star in Q2.1
(2009 June 20 -- 2009 July 20), but with a 2-d gap due to a safe mode
event. The star was, nevertheless, observed nearly continuously in Long Cadence
(LC) mode throughout the {\it Kepler} mission from Q0 to Q17 (2009 May 02
-- 2013 May 2013). A periodogram of the SC data shows the previously reported
rotational modulation, plus a weak higher frequency signal at 61.45~d$^{-1}$. 
The variability of KIC\,4768731 obtained from the {\it Kepler} light curves is
discussed in detail in Sect.~\ref{Sect_Pulsations}.

\begin{figure}
\includegraphics[width=\columnwidth]{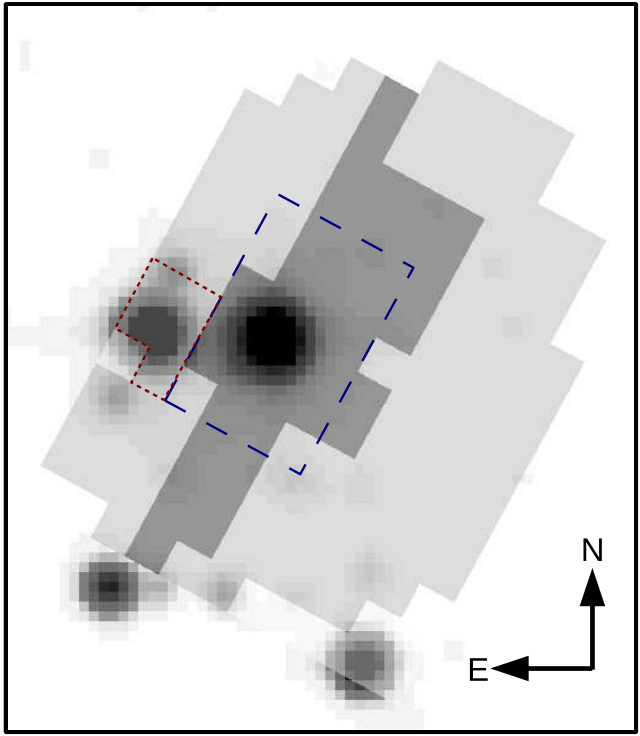}

\caption{{\it Kepler} pixel map for the SC data overlaid on the 2MASS $J$-band
image. The dark grey area contains the pixels used in the standard {\sc pdcsap}
extraction and in light grey are all the pixels downloaded from the spacecraft.
The areas highlighted in dashed and dotted lines are those used here to extract
light curves of KIC\,4768731 and KIC\,4768748, respectively.}

\label{PixelMap}
\end{figure}

The presence of the nearby companion star (KIC\,4768748) could potentially
affect the {\it Kepler} light curve of KIC\,4768731. Fig.~\ref{PixelMap} shows
the SC pixel map overlaid with the 2MASS $J$-band image. The standard {\sc
pdcsap} light curve extraction does not include the companion star, but it is
very close. Therefore,  using {\sc pyke} tools \citep{2012ascl.soft08004S}, we
extracted two light curves; one containing just the pixels centred on
KIC\,4768731 and the other centred on the companion star KIC\,4768748. The
light curve of KIC\,4768748 contains neither the rotational modulation nor the
high-frequency signal, while the light curve of KIC\,4768731 show the previously
reported rotational modulation plus the weak higher frequency signal.

\section{Frequency analysis of the {\it Kepler} data}
\label{Sect_Pulsations}

\begin{figure}
\centering
\includegraphics[width=\columnwidth]{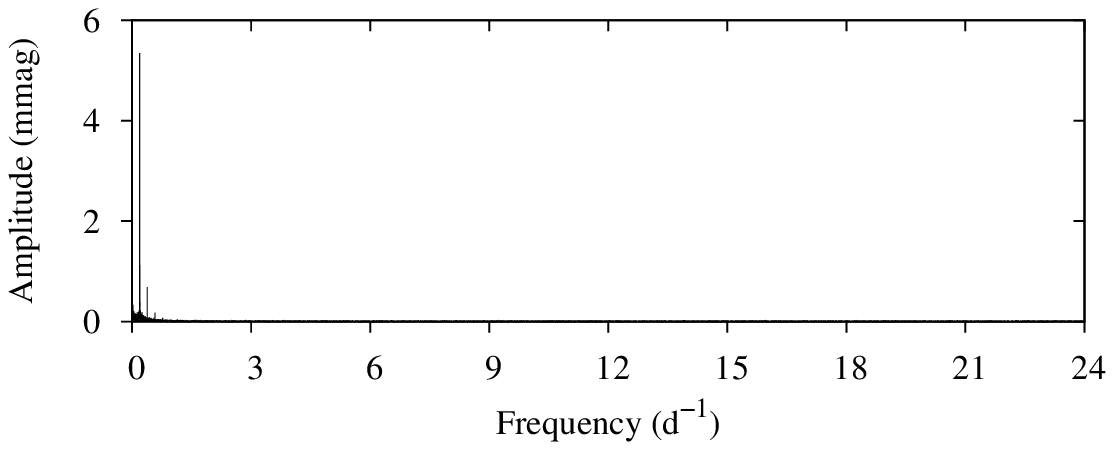}
\includegraphics[width=\columnwidth]{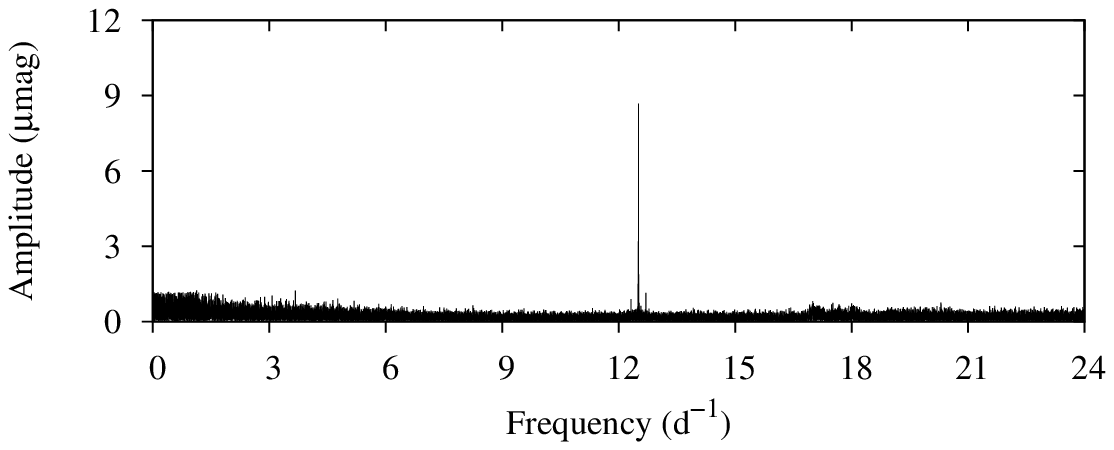}

\caption{Top panel: An amplitude spectrum for the Q1 to Q17 {\it Kepler} long
cadence data out to nearly the Nyquist frequency where only the low-frequency
rotational variations can be seen, as a consequence of the scale. Bottom panel:
An amplitude spectrum after a high-pass filter has removed the low-frequency
variation showing the Nyquist alias of the 61.45-d$^{-1}$ pulsation frequency
triplet. Note the three orders of magnitude change in the ordinate scale from
mmag in the top panel to $\umu$mag in the bottom panel.}

\label{fig:ft_all}
\end{figure}

KIC~4768731 was observed by {\it Kepler} in all quarters Q0--17 in LC, and
during one month, Q2.1, in SC. We use these data to study the pulsation
frequencies, amplitudes and phases for the full 4-yr data set. There is no
Nyquist limitation with {\it Kepler} LC data, although there is a reduction in
amplitude as a consequence of the pulsation periods being shorter than the
integration times. See \citet{2013MNRAS.430.2986M} for an explanation.
Fig.\,\ref{fig:ft_all} shows an amplitude spectrum of the Q0--17 data where only
the low-frequency peaks generated by the rotational variation can be seen in the
top panel. We ran a high-pass filter that removed the rotational variations from
the light curves, and the lower panel of Fig.\,\ref{fig:ft_all} shows the
amplitude spectrum of those filtered data. There is a clear frequency triplet
centred at 12.51\,d$^{-1}$. A closer look at these peaks shows them to be
Nyquist aliases of higher frequency peaks near 61.5\,d$^{-1}$, as is evident
from the SC data.

The rotation frequency found by fitting the highest peak in
Fig.~\ref{fig:ft_all} to the LC data by nonlinear least-squares gives $\nu_{\rm
rot} = 0.1919622 \pm 0.0000003$\,d$^{-1}$, or $P_{\rm rot} = 5.209358 \pm
0.000009$\,d.  

\begin{figure*}
\centering
\includegraphics[width=\columnwidth]{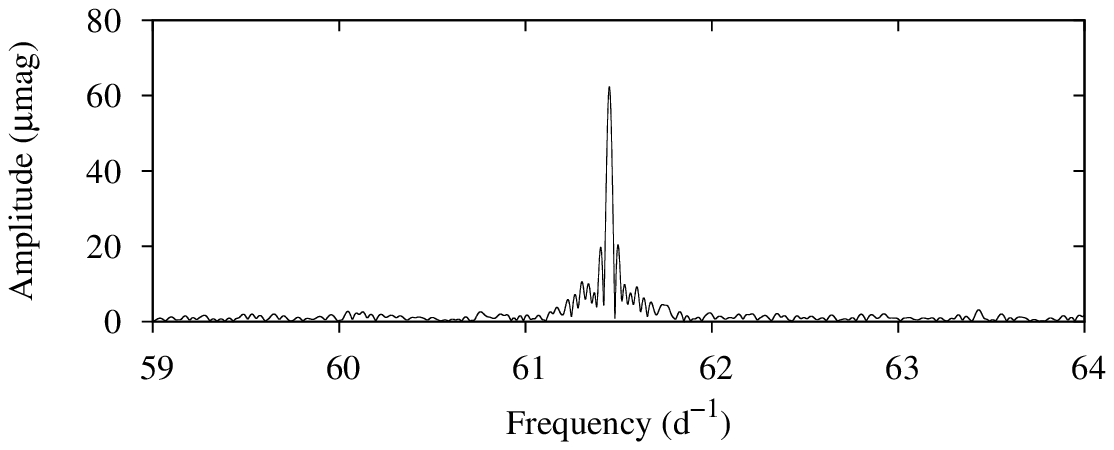}
\includegraphics[width=\columnwidth]{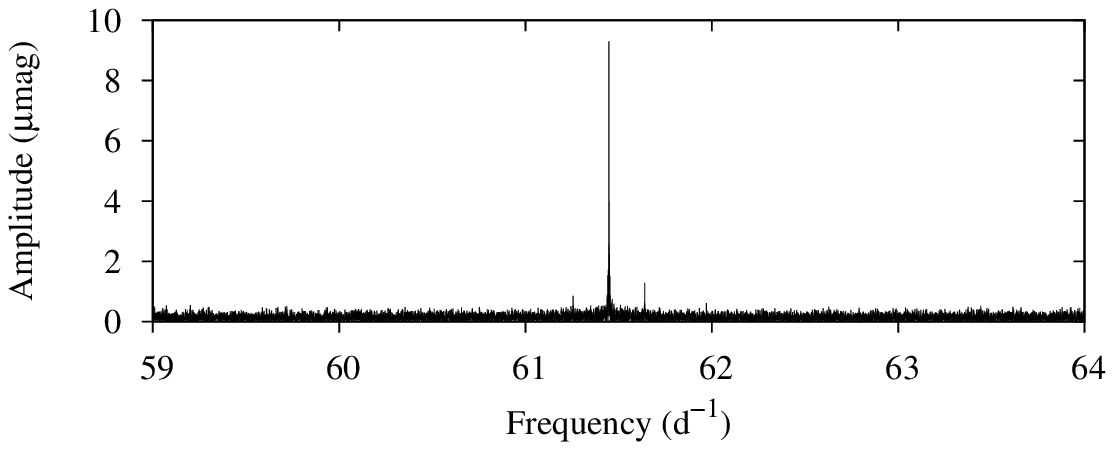}
\includegraphics[width=\columnwidth]{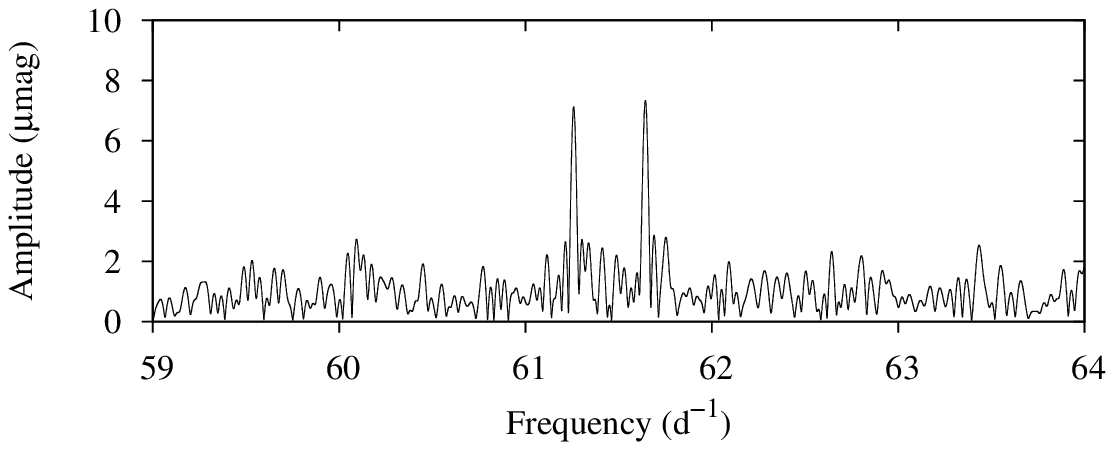}
\includegraphics[width=\columnwidth]{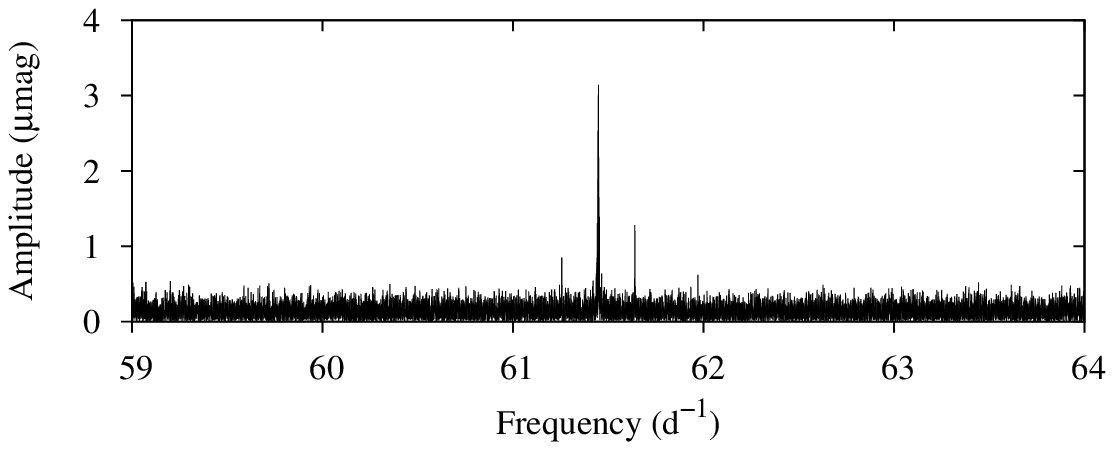}

\caption{Top left panel: An amplitude spectrum for the Q2.1 SC showing the
pulsation frequency near 61.45~d$^{-1}$. Bottom left panel: The SC data after
pre-whitening by $\nu_1$, where the rotational sidelobes are evident. Top right
panel: The same frequency range for the Q0--17 LC data where the oblique dipole
pulsation triplet can be seen. Bottom right panel: The LC data after
pre-whitening by $\nu_1$, where the rotational sidelobes are evident. The
residual amplitude around $\nu_1$ is the result of frequency or phase
variability of the pulsation frequency over the 4-yr time span of the data set,
as is seen in Fig.~\ref{fig:phamp}.}

\label{fig:ft_sc}
\end{figure*}

Fig.\,\ref{fig:ft_sc} shows an amplitude spectrum of the {\it Kepler} Q2.1 SC
data in the top left panel, where a clear peak is seen at 61.45\,d$^{-1}$. After
pre-whitening by the highest peak, the rotational sidelobes at $\nu_1  -
\nu_{\rm rot}$  and $\nu_1 + \nu_{\rm rot}$ become evident on the lower left
panel. This frequency triplet is equally spaced and is evidence of an oblique
dipole pulsation mode. This shows that the 12.51\,d$^{-1}$ peak seen in
Fig.\,\ref{fig:ft_all} is a Nyquist alias of this peak, since 61.45 -
$\nu_{\rm LC}$ = 12.51, where $\nu_{\rm LC} = 48.94$ is the {\it Kepler} LC
sampling frequency. The top right panel of Fig.\,\ref{fig:ft_sc} therefore
shows the same section of the amplitude spectrum for the entire Q0--17 {\it
Kepler} LC data set, where the frequency resolution is the highest available. Of
course, the amplitude is reduced because the integration times of the LC data
are longer than the pulsation period. From \citet{2012MNRAS.422..665M} we
calculate the reduction factor to be ${A}/{A_0} = \sin(\upi/n)/(\upi/n) = 5.5$,
where $n$ is the number of data points per oscillation cycle; this
explains the lower amplitude in the LC data when compared to the SC data. The
bottom right panel then shows the amplitude spectrum of the LC data after
prewhitening by $\nu_1$, where the rotational sidelobes can be seen, along with
residual amplitude around $\nu_1$. This residual amplitude is a consequence of
frequency variability over the 4-yr data set, as we show below.

\begin{table*}
\centering
\caption[]{A least-squares fit of the rotation frequency and the pulsational
dipole triplet for KIC~4768731  to the full 4-yr {\it Kepler} LC data set.
The zero point of the time scale is ${\rm BJD} 2455697.97056$, which coincides
with pulsation amplitude maximum and rotational light maximum. This is
typical for an oblique pulsator.}
\begin{tabular}{clrr}
\hline
\multicolumn{1}{c}{labels} &
\multicolumn{1}{c}{frequency} & \multicolumn{1}{c}{amplitude} &
\multicolumn{1}{c}{phase} \\
&\multicolumn{1}{c}{d$^{-1}$} & \multicolumn{1}{c}{$\umu$mag} &
\multicolumn{1}{c}{radians}  \\
\hline
\multicolumn{4}{c}{\it rotation}\\

$\nu_{\rm rot}$  & $0.1919622  \pm 0.0000003 $ & $5368.3  \pm 3.8 $ & $-3.0211  \pm 0.0007$ \\
[+2mm]
\multicolumn{4}{c}{\it pulsation -- long cadence (LC) data}\\
[+2mm]

$\nu_1 - \nu_{\rm rot}$  & $61.255837  \pm 0.000088 $ & $0.8  \pm 0.2 $ & $1.112  \pm 0.222$ \\
$\nu_1$  & $61.447800  \pm 0.000008 $ & $9.3  \pm 0.2 $ & $1.117  \pm 0.020$ \\
 $\nu_1 + \nu_{\rm rot}$ & $61.639762  \pm 0.000054 $ & $1.3  \pm 0.2 $ & $1.112  \pm 0.147$ \\
[+2mm]

\multicolumn{4}{c}{\it pulsation -- short cadence (SC) data}\\
[+2mm]

$\nu_1 - \nu_{\rm rot}$  & $61.255837  \pm 0.000088 $ & $6.5  \pm 0.7 $ & $2.842  \pm 0.115$ \\
$\nu_1$  & $61.447800  \pm 0.000008 $ & $62.2  \pm 0.7 $ & $3.057  \pm 0.012$ \\
 $\nu_1 + \nu_{\rm rot}$ & $61.639762  \pm 0.000054 $ & $8.0  \pm 0.7 $ & $2.842  \pm 0.092$ \\

\hline
\end{tabular}
\label{tab:puls}
\end{table*}

This triplet seen in Fig.\,\ref{fig:ft_sc} for both the SC and LC data is
equally split by exactly the rotation frequency to better than 1$\sigma$. This
is the signature of an oblique dipole pulsation mode. The three frequencies have
been fitted to the data by a combination of linear and nonlinear least-squares. Table\,\ref{tab:puls} shows the results where the frequency
splitting has been set to be exactly the rotation frequency.  A time zero point
was selected to set the phases of the sidelobes equal, and we found that the
phase of the central frequency is also equal, showing that the triplet is the
result of pure amplitude modulation. This is typical of pure oblique dipole
pulsation. We then find that the time of pulsation maximum, when all three
frequencies of the triplet have the same phase, $t_0 = 2455697.97056$, coincides
with maximum brightness of the rotational light modulation. That is shown
by the phase of the rotational frequency, near $-\upi$, with that time zero point.
Because we have fitted a cosine function, and the data are in magnitudes, the
phase at maximum rotational brightness is $\upi$ radians. This can also be seen
in the light curve, which we do not show here.

Maximum pulsation amplitude at rotational light minimum is expected for dipole
oblique pulsation where the spots that produce the rotational light variations
are aligned with the magnetic and pulsation poles. For KIC~4768731 the
relationship differs by $\upi$ radians from this simplest case, suggesting that
the spots, magnetic field pole and pulsation pole lie in the same plane, but are
not completely aligned. This is consistent with the improved oblique pulsator
model \citep{2011A&A...536A..73B}.

We can constrain the pulsation geometry within the oblique pulsator model
(\citealt{1982MNRAS.200..807K}; \citealt{2011A&A...536A..73B}) with the amplitudes of the
components of the frequency triplet:
\[
\tan i \tan \beta = (A_{+1} +A_{-1})/A_0 ,
\]
where $i$  is the rotational inclination, $\beta$ is the inclination
of the magnetic pole to the rotation axis, $A_{+1}$, $A_{-1}$ and $A_0$ are the
amplitudes of the rotational sidelobes and central frequency.
From this equation we obtain
$\tan i \tan \beta = 0.226 \pm 0.031$
and
$\tan i \tan \beta = 0.233 \pm 0.016$ from LC and SC data, respectively.
Using the stellar radius and $v \sin i$ obtained from spectral
analysis (Sect.~\ref{params}) and the rotation period, we obtain values of $i =
40^{+24}_{-10}$~{\degr} and $\beta = 15^{+9}_{-10}$~{\degr}. The angle between the
magnetic and rotation axes is therefore relatively small. In addition, $i + \beta < 90^\circ$ and only one pole is seen, hence the
simple, almost sinusoidal rotational light variations. 

\begin{figure}
\centering
\includegraphics[width=\columnwidth]{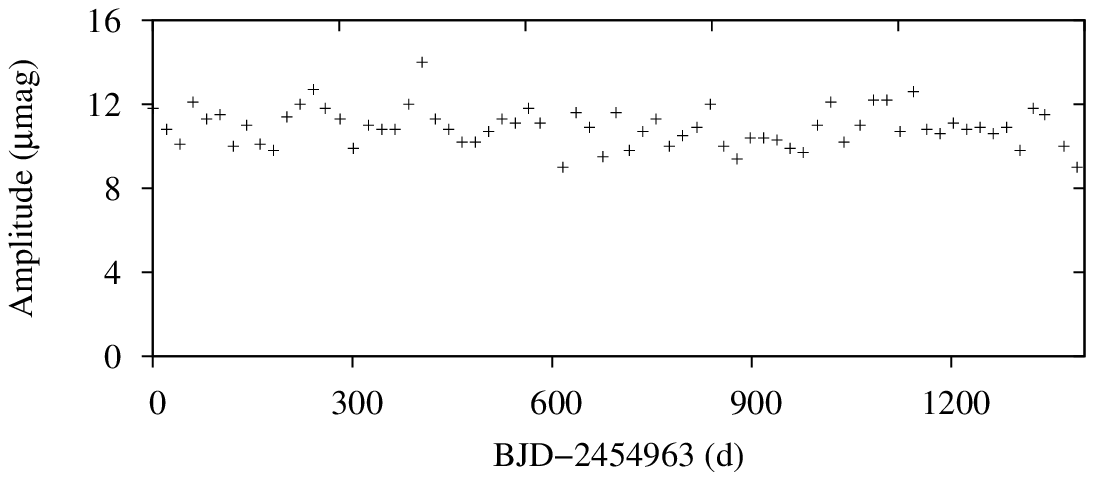}
\includegraphics[width=\columnwidth]{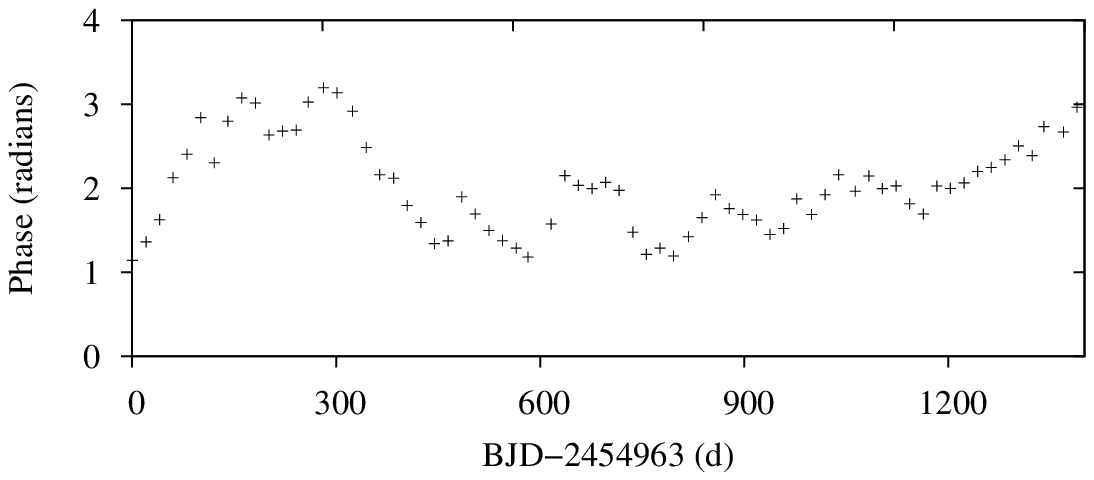}

\caption{Top panel: pulsation amplitude of $\nu_1 = 61.4478$\,d$^{-1}$ as a
function of time for 20-d segments of the Q0--17 KIC~4768731 data.  The
amplitude is stable, but may have a secular decrease over the 4 yr. Bottom
panel: pulsation phase for the same segments. There is clear variability in the
pulsation phase. There is an apparent oscillation with a period of about half
the {\it Kepler} orbital period, i.e. 186\,d, which may be instrumental in
origin. The larger, longer-term variations are most probably astrophysical in
origin.}

\label{fig:phamp}
\end{figure}

To study the residual amplitude around $\nu_1$ seen for the LC data, we fitted
$\nu_1$ to 20-d sections of the data by least-squares and examined the time
series of amplitude and phase. Fig.\,\ref{fig:phamp} shows that the amplitude is
stable over the 4-yr time span, and the phase is variable. Since phase and
frequency variability are indistinguishable, we interpret this as frequency
variability. This is the source of the residual amplitude in the amplitude
spectrum that is unresolved from $\nu_1$ as seen in the bottom right panel of
Fig.\,\ref{fig:ft_sc}. Such frequency variability is known in many other roAp
stars, as is discussed by \citet{2014MNRAS.443.2049H} for the roAp star
KIC\,7582608. The cause of such frequency variability is not yet known.

\section{Spectroscopic Observations}
\label{observations}

To examine the abundances in detail and to investigate line profile
variations as the star rotates, we obtained 33 high-resolution
spectra.

Two high-resolution ($R$=85\,000) spectra covering the wavelength range
3780--9000{\AA} were obtained using the fibre-fed High Efficiency and Resolution
Mercator Echelle Spectrograph \citep[HERMES;][]{2011A&A...526A..69R} mounted on
the 1.2-m Mercator Telescope at the Roque de los Muchachos Observatory, La
Palma. The data were reduced using the HERMES {\sc drs}\footnote{\url
{http://www.mercator.iac.es/instruments/hermes/hermesdrs.php}} pipeline
software.

A further two high-resolution spectra with a resolution $R$=67\,000 covering the
wavelength region 3660--7360{\AA} were obtained with the 2.5-m Nordic Optical
Telescope (NOT) using the FIbre-fed Echelle Spectrograph
\citep[FIES;][]{2014AN....335...41T}. These spectra were extracted with the
bespoke data reduction package, {\sc
fiestool}\footnote{\url{http://www.not.iac.es/instruments/fies/fiestool/}}.

A series of 26 high-resolution spectra were obtained in 2013 August using the
coud\'{e}-\'{e}chelle spectrograph attached to the 2.0-m telescope at the
Th\"{u}ringer Landessternwarte (TLS) Tautenburg, Germany. The spectrograph has a
resolving power of $R$ = 63\,000 and exposures cover the wavelength range from
4720--7360{\AA}. The spectra were reduced using standard ESO {\sc midas}
packages. All spectra were corrected in wavelength for individual instrumental
shifts by using a large number of telluric O$_2$ lines.

Table~\ref{Table_ObsLog} gives the dates of the observations, along with the
rotational phase ($\phi_{\rm rot}$) at the time of the midpoint of the exposures
and measured radial velocity (see Sect.~\ref{Sect_RV}). The rotational
phase is obtained from the fractional part of $(BJD - 2455697.97056) \times
0.1919622$, where phase zero is the time of maximum light.

A TLS spectrum of the companion star, KIC\,4768748, obtained on 2014 June
6, revealed that the star is indeed a late-K giant. In addition, the radial
velocity of this star is $-2.9 \pm 0.1$~km\,s$^{-1}$, compared to $-12.2 \pm
0.2$~km\,s$^{-1}$ for KIC\,4768731 (See Sect.~\ref{Sect_RV}). Thus, we conclude
that the two stars are not physically associated, KIC\,4768748 is a background
giant.

\begin{table*}
\begin{minipage}{137mm}
\caption{Log of spectral observations of KIC\,4768731. Date and BJD are given for the mid-point
of the exposure.}
\label{Table_ObsLog}
\begin{tabular}{lllllll} \hline
Instrument & Date        & Exp Time (s) & S/N &    BJD (TDB) & $\phi_{\rm rot}$ & RV (km\,s$^{-1}$) \\ \hline
HERMES     & 2011 Jul 13 22:53:56 & 1900 &  87 & 2455756.45765 & 0.227 & $-$12.0 \\ 
FIES       & 2013 Aug 04 00:43:06 & 2000 & 102 & 2456508.53370 & 0.597 & $-$12.3 \\ 
FIES       & 2013 Aug 06 00:19:54 & 1748 & 112 & 2456510.51759 & 0.978 & $-$11.8 \\ 
HERMES     & 2013 Aug 10 03:26:10 & 1800 &  80 & 2456514.64693 & 0.771 & $-$12.2 \\ 
TLS        & 2013 Aug 14 22:59:41 & 1800 &  80 & 2456519.46185 & 0.695 & $-$12.3 \\ 
TLS        & 2013 Aug 14 23:30:42 & 1800 &  73 & 2456519.48339 & 0.699 & $-$12.3 \\ 
TLS        & 2013 Aug 15 00:01:43 & 1800 &  70 & 2456519.50493 & 0.703 & $-$12.4 \\ 
TLS        & 2013 Aug 15 00:32:44 & 1800 &  63 & 2456519.52646 & 0.708 & $-$12.4 \\ 
TLS        & 2013 Aug 15 01:41:49 & 1800 &  58 & 2456519.57444 & 0.717 & $-$12.3 \\ 
TLS        & 2013 Aug 15 02:12:50 & 1800 &  49 & 2456519.59598 & 0.721 & $-$12.3 \\ 
TLS        & 2013 Aug 15 02:43:51 & 1800 &  46 & 2456519.61752 & 0.725 & $-$11.7 \\ 
TLS        & 2013 Aug 15 21:54:18 & 2400 & 108 & 2456520.41643 & 0.879 & $-$12.1 \\ 
TLS        & 2013 Aug 16 01:58:32 & 2400 &  85 & 2456520.58604 & 0.911 & $-$12.7 \\ 
TLS        & 2013 Aug 16 21:44:06 & 2400 &  95 & 2456521.40934 & 0.069 & $-$12.4 \\ 
TLS        & 2013 Aug 17 01:59:27 & 2400 &  71 & 2456521.58667 & 0.103 & $-$12.4 \\ 
TLS        & 2013 Aug 17 22:59:50 & 2400 &  89 & 2456522.46192 & 0.271 & $-$12.3 \\ 
TLS        & 2013 Aug 21 19:54:11 & 2400 &  76 & 2456526.33295 & 0.014 & $-$11.9 \\ 
TLS        & 2013 Aug 22 20:10:07 & 2400 &  72 & 2456527.34400 & 0.208 & $-$11.9 \\ 
TLS        & 2013 Aug 23 19:51:48 & 2400 &  76 & 2456528.33127 & 0.398 & $-$12.2 \\ 
TLS        & 2013 Aug 24 00:50:49 & 2417 &  70 & 2456528.53891 & 0.438 & $-$12.2 \\ 
TLS        & 2013 Aug 26 20:58:21 & 2400 &  67 & 2456531.37743 & 0.983 & $-$12.4 \\ 
TLS        & 2013 Aug 26 22:25:29 & 2400 &  78 & 2456531.43794 & 0.994 & $-$12.4 \\ 
TLS        & 2013 Aug 26 23:06:30 & 2400 &  73 & 2456531.46642 & 1.000 & $-$12.3 \\ 
TLS        & 2013 Aug 27 01:42:35 & 2400 &  60 & 2456531.57481 & 0.021 & $-$12.3 \\ 
TLS        & 2013 Aug 27 20:42:41 & 2400 &  45 & 2456532.36653 & 0.172 & $-$12.1 \\ 
TLS        & 2013 Aug 27 21:24:35 & 2400 &  98 & 2456532.39563 & 0.178 & $-$12.2 \\
TLS        & 2013 Aug 27 22:07:30 & 2400 &  95 & 2456532.42543 & 0.184 & $-$12.3 \\ 
TLS        & 2013 Aug 27 22:48:31 & 2400 &  84 & 2456532.45392 & 0.189 & $-$12.3 \\ 
TLS        & 2013 Aug 27 23:29:32 & 2400 &  90 & 2456532.48240 & 0.195 & $-$12.4 \\ 
TLS        & 2013 Aug 28 00:10:33 & 2400 &  76 & 2456532.51088 & 0.200 & $-$12.3 \\ 
TLS        & 2014 Jun 17 23:10:15 & 2400 &  86 & 2456826.46545 & 0.628 & $-$12.4 \\
TLS        & 2015 May 10 22:55:57 & 2400 &  84 & 2457153.45487 & 0.398 & $-$12.2 \\
TLS        & 2015 May 11 01:07:39 & 2400 &  75 & 2457153.54733 & 0.416 & $-$12.1 \\
\hline
\end{tabular} \newline
Notes: The rotational phase ($\phi_{\rm rot}$) is given by the fractional
part of $(BJD - 2455697.97056) \times 0.1919622$,
where phase zero is the time of maximum light. 
The signal-to-noise (S/N) ratio obtained using the {\sc DER\_SNR} algorithm \citep{2008ASPC..394..505S}.
\end{minipage}
\end{table*}

\section{Spectroscopic Analysis}
\label{params}

A detailed spectroscopic analysis of the 2011 HERMES spectrum was presented in
\cite{2015MNRAS.450.2764N} who identified KIC\,4768731 as an Ap\,CrSrEu star.
Table~\ref{Table_Params} summarizes the basic stellar parameters obtained for
this star.

\begin{table}
\caption{Basic stellar parameters of KIC\,4768731.}
\label{Table_Params}
\begin{tabular}{ccc} \hline
Parameter     & Value & Units \\ \hline
$T_{\rm eff}$ & 8100 $\pm$ 200     & K \\
$\log g$      & 4.0 $\pm$ 0.2      & \\
{[Fe/H]}      & +0.31 $\pm$ 0.24   & \\
$V_{\rm micro}$ & 0.5 $\pm$ 0.3    & km\,s$^{-1}$ \\
$v \sin i$    & 14.8 $\pm$ 1.6     & km\,s$^{-1}$ \\
$M$           & 2.11 $\pm$ 0.27    & M$_{\sun}$ \\
$R$           & 2.39 $\pm$ 0.68    & R$_{\sun}$ \\
$\log L$      & 1.34 $\pm$ 0.25    & L$_{\sun}$ \\
$M_V$         & 1.43 $\pm$ 0.67    & mag. \\                                                        
Sp. Type      &   A5\,Vp~SrCr(Eu)  & \\ \hline
\\
\end{tabular}
\newline {\bf Note:} Mass ($M$) and radius ($R$) are estimated using the
\cite{2010A&ARv..18...67T} calibration.
\end{table}

\subsection{Magnetic field}

The ratio of the strengths of the Fe\,{\sc ii} 6147.7{\AA} and 6149.2{\AA} lines
can be used to estimate the mean magnetic field modulus $\langle H \rangle$
\citep{1992A&A...256..169M}. From the individual spectra we obtain a mean value
for the relative intensification of the Fe\,{\sc ii} 6147.7{\AA} line with
respect to the Fe\,{\sc ii} 6149.2{\AA} line, $\Delta W_\lambda/\overline{W}_\lambda$, of
0.08 $\pm$ 0.03, where $\Delta W_\lambda = W_\lambda(6147.7) -
W_\lambda(6149.2)$. This yields a mean magnetic field modulus, $\langle H
\rangle$, of 2.7 $\pm$ 0.8~kG using the empirical relationship given in
\cite{1992A&A...256..169M}. The error bar includes both the scatter in the
observed values of $\Delta W_\lambda/\overline{W}_\lambda$ and the uncertainty
in the calibration for $\langle H \rangle$. The individual values for $\Delta
W_\lambda/\overline{W}_\lambda$ show no significant variation with rotational
phase.

For the magnetic modulus obtained above the splitting of the 6149.2{\AA} Zeeman
doublet components would be $\sim$0.13~{\AA} \citep{1997A&AS..123..353M}.
However, there is no sign of any splitting in our spectra due to the relatively
large $v \sin i$ of this star.

\subsection{Radial velocity variations?}
\label{Sect_RV}

\begin{figure}
\includegraphics[width=\columnwidth]{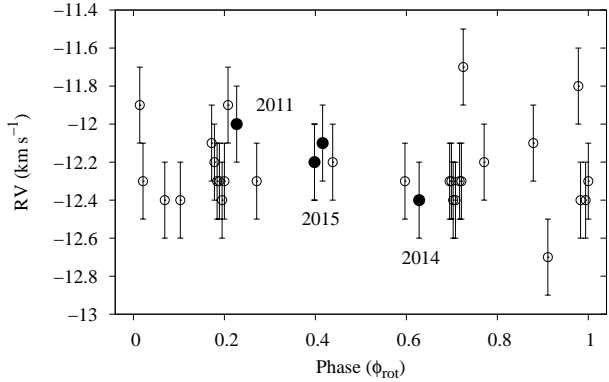}
\caption{Radial velocity measurements show no trend with rotational phase.
The filled circles indicate the values obtained in 2011, 2014 and 2015 and
show that they are not significantly different from those from 2013 (open circles).}
\label{RV_phase}
\end{figure}

A radial velocity (RV) for each spectrum was determined by cross-correlation
with a synthetic spectrum covering the wavelength range 5000--5800{\AA} using
the parameters and abundances obtained from the detailed spectroscopic analysis.
The barycentric radial velocities are given in Table~\ref{Table_ObsLog} and the
formal uncertainties are 0.2~km\,s$^{-1}$. The average radial velocity from the
spectra obtained in 2013 is $-$12.2 $\pm$ 0.2~km\,s$^{-1}$. However,
\cite{1990A&AS...83...91F} reported a value of $-$29$\pm$1.2~km\,s$^{-1}$ based
on four observations obtained on 1984 June 27, 1986 August 2, 30 and 31. They
used the Fehrenbach Objective Prisms on the Schmidt telescope at the
Observatorie de Haute-Provence and the external error is given as
3.5~km\,s$^{-1}$. The difference between the two epochs of $\sim$17~km\,s$^{-1}$
suggests that the star could be a spectroscopic binary. The HERMES spectrum
taken in July 2011 has a radial velocity that is consistent with those taken
during August 2013. In June 2014 and May 2015 we obtained further TLS spectra
and their RV values are consistent with the previous years
(Fig.~\ref{RV_phase}). Hence, we find no significant change in RV over a period
of 4\,yr. 

\subsection{Spectral line profile variations}
\label{lpv}

\begin{figure}
\includegraphics[width=\columnwidth]{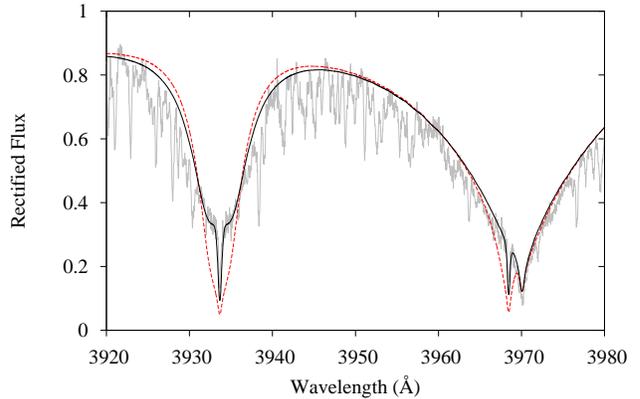}
\caption{The Ca H and K lines in the FIES spectrum (grey line) taken on 2013
August 6 shows the distinctive pointed core due to element
stratification. The solid line shows a synthetic line profile created using a step-shaped stratified abundance
profile, while the red dashed-line shows a normal non-stratified synthesis.} 
\label{Fig_Ca_HK}
\end{figure}

The individual spectra were examined for evidence of spectral line profile
variations with rotational phase. The H$\alpha$ and H$\beta$ Balmer lines and
the Na\,D lines do not show any variation with phase. Similarly, the HERMES and
FIES spectra of the Ca H and K lines do not show any variation with rotational
phase, but have the unusual line profiles associated with abundance
stratification in the atmosphere
\citep{1994A&A...283..189B,2002A&A...384..545R}. The sharp cores of Ca H and K
lines can be recreated using a step-shaped stratified abundance profile with
[Ca/H] = +1.0~dex for layers deeper than $\log(\tau_{5000}) = -1.0$ and [Ca/H] =
$-$2~dex for higher layers (Fig.~\ref{Fig_Ca_HK}). In contrast, a global
enhancement of [Ca/H] = +0.5~dex is unable to reproduce the observed line
profile.

\begin{figure}
\includegraphics[width=\columnwidth]{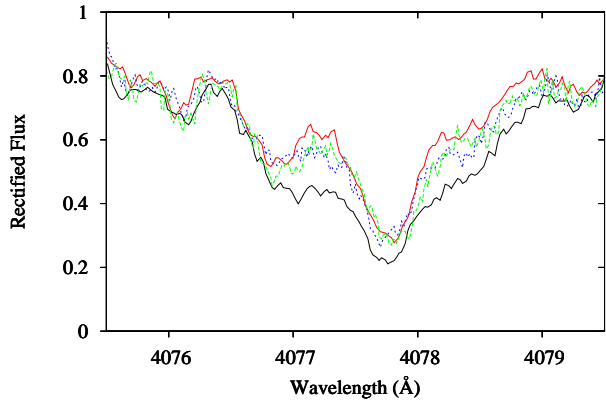}
\caption{Variations in the profile shape of the Sr\,{\sc ii} 4077.7{\AA} line with
rotational modulation phase. The solid black line was taken at $\phi_{\rm rot}$
= 0.978, the solid red line at $\phi_{\rm rot}$ = 0.597. The dashed green and
dotted blue lines were taken at $\phi$ = 0.771 and $\phi_{\rm rot}$ = 0.227,
respectively.}
\label{Fig_Sr}
\end{figure}

The HERMES and FIES spectra of the Sr\,{\sc ii} 4077.7{\AA} line shows
considerable variation with rotational phase (Fig.~\ref{Fig_Sr}). The line is
strongest at $\phi_{\rm rot}$ = 0.978 and weakest at $\phi_{\rm rot}$ = 0.597.
This is consistent with the weakest line strength occurring at or around
$\phi_{\rm rot}$ = 0.5. The other two spectra are very similar, but their phases
with respect to $\phi_{\rm rot}$ = 0.5 are $-$0.271 and +0.273, respectively.
This suggests that there is a region of enhanced Sr crossing the stellar
meridian at the time of maximum light ($\phi_{\rm rot}$ = 0.0).

\begin{figure}
\includegraphics[width=\columnwidth]{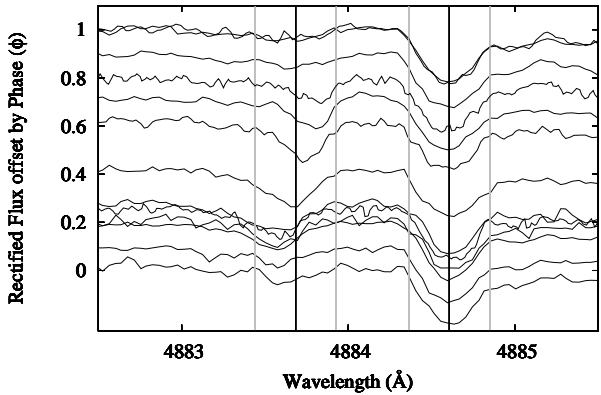}

\caption{Changes in line profile strength and shape with rotational phase for
the Y\,{\sc ii} line at 4883.7{\AA}. The rectified spectra have been
offset so that their continuum levels correspond to their
rotational phases. The nearby Cr\,{\sc ii} line at
4884.6{\AA} shows little shape variation with phase. The vertical black and grey
lines indicate the line centres and the $\pm$15~km\,s$^{-1}$ $v \sin i$ widths,
respectively.}

\label{Fig_Y}
\end{figure}

Fig.~\ref{Fig_Y} shows the variation of the Y\,{\sc ii} 4883.7{\AA} line
profile with rotational phase. There is a clear absorption feature moving
redward with increasing phase, first appearing on the blue edge of the line
profile at around $\phi_{\rm rot}$ = 0.1 and disappearing around $\phi_{\rm
rot}$ = 0.9. The feature crosses the centre of the line around $\phi_{\rm rot}$
= 0.5, coinciding with the time of minimum light in the rotation cycle.
Similar behaviour is seen in the Y\,{\sc ii} 4900.1 and Ba\,{\sc ii}
6141.7{\AA} lines. This is in anti-phase to the variation of the Sr\,{\sc ii}
line.

\begin{table*}
\begin{minipage}{155mm}
\caption{Elemental abundances obtained for KIC\,4768731 from TLS spectra.
Abundances are given in the form $\log A({\rm El}) = \log(N_{\rm El}/N_{\rm H}) + 12$ obtained from spectra taken on six different nights and rotational phases ($\phi_{\rm rot}$).
$n$ is the maximum number of lines for an element used in the abundance averages, which can
be slightly fewer at different rotational phases. Line-to-line standard deviation error bars are given when $n>2$.
In column 9 the Solar abundances from \citet{2009ARA&A..47..481A} are given in for reference.}
\label{table_abunds}
\begin{tabular}{lllllllll} \hline
El&$n$& Aug 14--15      & Aug 15--16      & Aug 16--17      & Aug 23--24      & Aug 26--27      & Aug 27--28      & Solar \\
  &   &($\phi_{\rm rot} = 0.71$)&($\phi_{\rm rot} = 0.89$)&($\phi_{\rm rot} = 0.09$)&($\phi_{\rm rot} = 0.42$)&($\phi_{\rm rot} = 1.00$)&($\phi_{\rm rot} = 0.19$)\\
\hline
 C& 19&8.45 $\pm$ 0.33 & 8.26 $\pm$ 0.27 & 8.47 $\pm$ 0.36 & 8.51 $\pm$ 0.28 & 8.43 $\pm$ 0.34 & 8.46 $\pm$ 0.37 & 8.43 \\
 N&  1&8.16            & 8.38            & 8.06            & 8.20            & 7.94            & 8.15            & 7.83 \\
 O&  5&9.37 $\pm$ 0.22 & 9.10 $\pm$ 0.51 & 9.14 $\pm$ 0.41 & 8.62 $\pm$ 0.55 & 9.16 $\pm$ 0.42 & 9.40 $\pm$ 0.21 & 8.69 \\
Na&  3&7.13 $\pm$ 0.06 & 7.08            & 7.21 $\pm$ 0.12 & 6.81 $\pm$ 0.31 & 7.22 $\pm$ 0.12 & 6.77            & 6.24 \\
Mg&  8&7.83 $\pm$ 0.18 & 7.87 $\pm$ 0.21 & 7.82 $\pm$ 0.20 & 7.73 $\pm$ 0.17 & 7.86 $\pm$ 0.20 & 7.84 $\pm$ 0.16 & 7.60 \\
Al&  2&7.47            & 7.43            & 7.45            & 7.04            & 7.34            & 7.15            & 6.45 \\
Si& 34&7.53 $\pm$ 0.27 & 7.66 $\pm$ 0.24 & 7.75 $\pm$ 0.21 & 7.45 $\pm$ 0.26 & 7.74 $\pm$ 0.23 & 7.56 $\pm$ 0.22 & 7.51 \\
 S&  9&7.34 $\pm$ 0.43 & 7.56 $\pm$ 0.49 & 7.57 $\pm$ 0.25 & 7.02 $\pm$ 0.26 & 7.27 $\pm$ 0.33 & 7.22 $\pm$ 0.34 & 7.12 \\
Ca& 18&6.81 $\pm$ 0.21 & 6.77 $\pm$ 0.20 & 6.88 $\pm$ 0.21 & 6.80 $\pm$ 0.18 & 6.81 $\pm$ 0.20 & 6.83 $\pm$ 0.22 & 6.34 \\
Sc&  7&2.45 $\pm$ 0.10 & 2.47 $\pm$ 0.35 & 2.46 $\pm$ 0.33 & 2.66 $\pm$ 0.45 & 2.30 $\pm$ 0.35 & 2.53 $\pm$ 0.45 & 3.15 \\
Ti& 42&5.08 $\pm$ 0.29 & 5.00 $\pm$ 0.28 & 5.13 $\pm$ 0.25 & 5.12 $\pm$ 0.24 & 5.17 $\pm$ 0.24 & 5.19 $\pm$ 0.27 & 4.95 \\
 V&  8&5.10 $\pm$ 0.30 & 5.17 $\pm$ 0.30 & 5.19 $\pm$ 0.28 & 5.14 $\pm$ 0.30 & 5.22 $\pm$ 0.28 & 5.21 $\pm$ 0.24 & 3.93 \\
Cr&111&7.36 $\pm$ 0.18 & 7.50 $\pm$ 0.18 & 7.51 $\pm$ 0.17 & 7.21 $\pm$ 0.16 & 7.55 $\pm$ 0.17 & 7.38 $\pm$ 0.19 & 5.64 \\
Mn& 17&5.77 $\pm$ 0.35 & 5.82 $\pm$ 0.36 & 5.72 $\pm$ 0.32 & 5.79 $\pm$ 0.28 & 5.84 $\pm$ 0.34 & 5.69 $\pm$ 0.30 & 5.43 \\
Fe&142&7.79 $\pm$ 0.18 & 7.83 $\pm$ 0.18 & 7.83 $\pm$ 0.17 & 7.80 $\pm$ 0.19 & 7.81 $\pm$ 0.18 & 7.83 $\pm$ 0.18 & 7.50 \\
Co&  6&5.78 $\pm$ 0.25 & 6.11 $\pm$ 0.37 & 6.14 $\pm$ 0.33 & 5.84 $\pm$ 0.24 & 6.28 $\pm$ 0.26 & 5.74 $\pm$ 0.18 & 4.99 \\
Ni& 31&6.30 $\pm$ 0.34 & 6.37 $\pm$ 0.36 & 6.33 $\pm$ 0.37 & 6.18 $\pm$ 0.28 & 6.24 $\pm$ 0.35 & 6.39 $\pm$ 0.31 & 6.22 \\
Cu&  1&3.09            & 3.05            & 3.12            & 3.34            & 3.63            & 3.08            & 4.19 \\
Zn&  1&4.73            & 4.70            & 4.53            & 4.68            & 4.69            & 4.72            & 4.56 \\
 Y&  9&3.05 $\pm$ 0.24 & 2.74 $\pm$ 0.39 & 2.90 $\pm$ 0.25 & 3.12 $\pm$ 0.21 & 2.82 $\pm$ 0.20 & 2.79 $\pm$ 0.35 & 2.21 \\
Zr&  3&3.60            & 4.67 $\pm$ 0.77 & 4.36 $\pm$ 0.69 & 3.49            & 3.48            & 4.27 $\pm$ 0.58 & 2.58 \\
Ba&  2&2.50            & 1.81            & 1.87            & 2.52            & 1.80            & 2.05            & 2.18 \\
La&  2&1.60            & 1.34            & 1.91            & 0.77            & 1.23            & 1.75            & 1.10 \\
Nd&  6&2.34 $\pm$ 0.15 & 2.24 $\pm$ 0.02 & 2.31 $\pm$ 0.14 & 2.19 $\pm$ 0.24 & 2.40 $\pm$ 0.02 & 2.28 $\pm$ 0.06 & 1.42 \\
Eu&  3&2.65 $\pm$ 0.31 & 2.50 $\pm$ 0.11 & 2.33 $\pm$ 0.36 & 1.93 $\pm$ 0.50 & 2.47 $\pm$ 0.34 & 2.20 $\pm$ 0.26 & 0.52 \\
\hline
\end{tabular}         
\end{minipage}
\end{table*}

\begin{figure}
\includegraphics[width=\columnwidth]{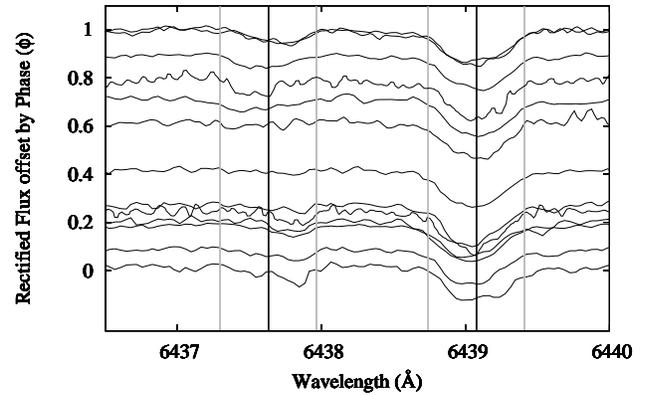}

\caption{Changes in line profile strength and shape with rotational phase for
the Eu\,{\sc ii} line at 6437.6{\AA}. The rectified spectra have been
offset so that their continuum levels correspond to their
rotational phases. The nearby Ca\,{\sc i} line at
6439.1{\AA} shows little shape variation with phase. The vertical black and grey lines
indicate the line centres and the $\pm$15~km\,s$^{-1}$ $v \sin i$ widths,
respectively.}

\label{Fig_Eu}
\end{figure}

Fig.~\ref{Fig_Eu} shows the variation of the Eu\,{\sc ii} 6437.6{\AA}
line profile with rotational phase. The line is quite weak around
$\phi_{\rm rot}$ = 0.4. By $\phi_{\rm rot}$ = 0.7 an absorption feature can be
seen in the blue wing of the profile. As phase increases the feature moves
redward across the line profile, reaching a maximum around $\phi_{\rm rot}$ =
0.0 when it is close to the line centre. The feature then moves to the red wing
and weakens, disappearing around $\phi_{\rm rot}$ = 0.2. Similar behaviour is
seen in other Eu lines. The variations in the Eu lines are in anti-phase with
those for Y and Ba lines, but in phase with the Sr line.

To further investigate line strength variations with rotational phase, we
selected six nights in 2013 August with multiple TLS spectra. For each night the
spectra were co-added to produce a single nightly spectrum with signal-to-noise
(S/N) ratios in the range 100 -- 150. For each of the six spectra abundances
were obtained using the same spectral synthesis methodology as
\cite{2015MNRAS.450.2764N}. The results of the spectral fitting are given in
Table~\ref{table_abunds}. The overall abundance pattern is consistent with that
found by \cite{2015MNRAS.450.2764N} and with the spectral classification
presented in Sect.~\ref{SpecClass}: Cr is considerably enhanced, Eu is modestly
enhanced, while Ca and Si are relatively normal. V and Co are also considerably
enhanced and there is general modest ($\sim +0.3$~dex) overabundance of most
other transition-group elements, with the exception of Sc, which is depleted by
over 0.5~dex. Several elements, Mg, Si, Cr, Co and Eu, have abundance variations
in phase with rotation -- they are strongest when the star is the brightest. In
addition, the same behaviour was noted for the Sr line above. The Y and Ba
lines, on the other hand, vary in anti-phase with rotation -- they are
strongest when the star is the faintest. Other elements, including Ca,
Ti, V, Mn and Fe, show little variation with phase or the results are
inconclusive.

To search for the presence of other rare earth elements, the
individual TLS spectra from 2013 were co-added to produce a single spectrum with
a S/N of 245. This spectrum was visually searched for lines due to various rare earth
elements. With the exception of the already detected La, Nd and Eu, only
tentative detections or upper limits were obtained (see
Table~\ref{rare_earths}).

\begin{table} 
\caption{Rare earth elements in KIC\,4768731.
Abundances are given in the form $\log A({\rm El}) = \log(N_{\rm El}/N_{\rm H}) + 12$.
A colon (:) indicates a tentative detection, where other lines of the
element have upper limits below the given value.
$n$ is the number of lines used to obtain the abundance averages, except
in the cases of upper limits or tentative detections where it is the number of
lines searched.
In column 4 the Solar abundances from \citet{2009ARA&A..47..481A} are
given in for reference.}
\label{rare_earths}
\begin{tabular}{llll} \hline
El & n & $\log A({\rm El})$ & Solar \\ \hline
La & 2 & 1.40               & 1.10 \\
Ce & 6 & $<$2.5             & 1.58 \\
Pr & 3 & 3.2:               & 0.72 \\
Nd & 6 & 2.25 $\pm$ 0.12    & 1.42 \\
Sm & 2 & 2.2:               & 0.96 \\
Eu & 3 & 2.45 $\pm$ 0.11    & 0.52 \\
Gd & 4 & 3.0:               & 1.07 \\
Tb & 3 & $<$2.5             & 0.30 \\
Dy & 4 & $<$3.3             & 1.10 \\
Er & 2 & 3.5:               & 0.92 \\
Tm & 2 & $<$2.5             & 0.10 \\
Yb & 6 & $<$3.5             & 0.84 \\
Lu & 5 & $<$2.0             & 0.10 \\
\hline
\end{tabular}
\end{table}

The relatively large, and variable, scatter in the standard deviations of some
of the abundances obtained is probably the result of the surface inhomogeneities
and vertical stratification present in the atmosphere of this Ap star. A
detailed spectral analysis taking into account surface inhomogeneities and
abundance stratification will be presented in Niemczura, Shulyak et al. (in
preparation).

\section{Non-adiabatic oscillation modelling}

\label{sec:model}

The mechanism responsible for exciting the oscillations observed in roAp stars
is still not fully understood. The opacity mechanism acting on the hydrogen
ionization region leads to the excitation of high radial-order  acoustic
pulsations in models of cool Ap stars with fully radiative envelopes, where
convection is assumed to be suppressed by the strong magnetic field
\citep{2001MNRAS.323..362B}. This mechanism has been shown to provide a promising
explanation for the observed pulsations in roAp stars with pulsation frequencies
below the acoustic cutoff \citep{2002MNRAS.333...47C}. However, in a fraction of the known
roAp pulsators, the observed frequencies are too high to be driven in this way
\citep{2013MNRAS.436.1639C} and an alternative mechanism needs to be considered, such as
the effect of turbulent pressure suggested by \cite{2013MNRAS.436.1639C} and recently
found to be a likely explanation for pulsations in an Am star
\citep{2014ApJ...796..118A}.  

\begin{figure}
\includegraphics[width=\columnwidth]{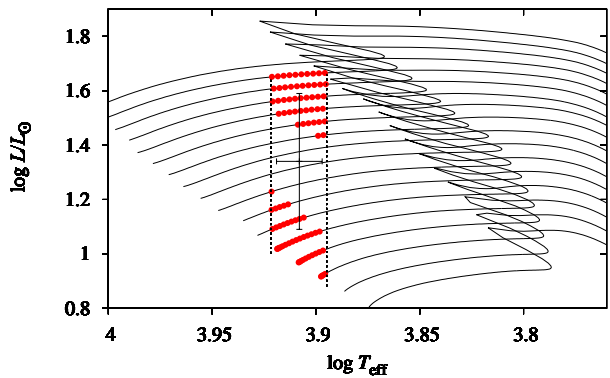}

\caption{HR diagram showing the results of the non-adiabatic oscillation
modelling. The cross indicates the location and 1$\sigma$ uncertainty of KIC\,4768731. The
solid lines are the evolutionary tracks (starting from the ZAMS) covering masses
from 1.6~M$_{\sun}$ (bottom) to 2.4 (top) in steps of 0.05~M$_{\sun}$. The
vertical dashed-lines indicate the range in temperatures searched. Models found
to be unstable at the observed frequency are marked by filled-circles. In the
lower luminosity region, the modes excited are of low radial-order, while in the
higher luminosity region they are of high radial-order.}

\label{fig_eq_po_t}
\end{figure}

The oscillation frequency observed in KIC~4768731 is below the acoustic cutoff
frequency derived from models covering the region of the HR diagram where the
star is located.  To check if the opacity mechanism can drive the observed
pulsation, and further constrain the global parameters of the star, we carried
out a linear, non-adiabatic stability analysis of a grid of models covering that
region of the HR diagram. Evolutionary tracks were produced with the {\sc mesa}
stellar evolution code \citep{2013ApJS..208....4P} for stellar masses in
the range 1.6--2.4~M$_{\sun}$, in intervals of 0.05~M$_{\sun}$, and with
initial mass fraction of hydrogen and helium of $X=0.70$ and $Y=0.28$,
respectively. Mixing beyond the convective core during the main-sequence has
been considered, taking the convective overshoot parameter defined in
\cite{2013ApJS..208....4P} (their equation~(9)) to be $f_{\rm ov}=0.016$. Only
models with $T_{\rm eff}$ in the range 7850--8350~K were considered for the
stability analysis (vertical dashed lines in Fig.~\ref{fig_eq_po_t}). The mass,
luminosity, and effective temperature extracted from the {\sc mesa} models were
used to generate the equilibrium models necessary for the non-adiabatic
computations. 

The non-adiabatic analysis followed closely that of \cite{2013MNRAS.436.1639C}, except
that we have considered only their `polar'  models (models with fully radiative
envelopes) as these are the models in which roAp-type pulsations are found to be
driven by the opacity mechanism. Following that work, for each set of mass,
luminosity, and effective temperature, we have considered four different case
studies. In the first (or {\it  standard} ) case the equilibrium model has a
surface helium abundance of $Y_{\rm sur}=0.01$ and an atmosphere that extends to
a minimum optical depth of $\tau_{\rm min}=3.5\times10^{-5}$.  Moreover, in the
pulsation analysis a fully reflective boundary condition is applied at the
surface for this case study. The other three case studies are obtained by
swapping these properties, one at the time to: $Y_{\rm sur}=0.1$,  $\tau_{\rm
min}=3.5\times10^{-4}$, and a transmissive boundary condition.  Together, these
four case studies cover the main uncertainties in the modelling.  If modes with
frequencies similar to that observed in KIC\,4768731 are found unstable in at
least one of the case studies, then the corresponding model parameter set is
identified as unstable at the observed frequency. 

Models found to be unstable at the observed frequency are marked by filled-circles in
Fig.~\ref{fig_eq_po_t}. Inspection of this figure shows that there are two
independent regions within the part of the HR diagram explored in which
pulsations with the observed frequency are predicted to be excited.  In the
lower luminosity region, the modes excited at the observed frequency are of low
radial-order, characteristic of $\delta$-Scuti stars, while in the higher
luminosity region they are of high radial-order, characteristic of roAp stars. 

The non-adiabatic analysis performed here does not include the direct effect of
the magnetic field on pulsations, but only its indirect effect through the
suppression of convection in the stellar envelope. However, it is known from
studies addressing the direct effect of the magnetic field that the coupling of
acoustic and magnetic waves in the outer layers of roAp stars leads to wave
energy losses through the dissipation of acoustic waves in the
atmosphere~\citep{2008MNRAS.386..531S} and magnetic waves in the
interior~\citep{2000MNRAS.319.1020C}. In fact,  \cite{2005MNRAS.360.1022S} has
shown that energy losses through the latter process are particularly significant
for low radial-order modes. These modes are then stabilised, explaining their
absence in roAp stars.  If this is the case for KIC\,4768731, then the lower
luminosity unstable region shown in Fig.~\ref{fig_eq_po_t} is likely to be
spurious, resulting from the non-inclusion of the direct effect of the magnetic
field on pulsations in the non-adiabatic computations. Our non-adiabatic
calculations thus point to KIC\,4768731 being a relatively evolved and luminous
star.

\section{Discussion and conclusions}

Given the identification of KIC\,4768731 as an Ap star, the pulsation is almost
certainly of the roAp type, and with a frequency of 61.45\,d$^{-1}$
(corresponding to a period of 23.43\,min) this star only just misses out on
being the longest period roAp star known (HD\,177765;
\citealt{2012MNRAS.421L..82A}, 23.56\,min). Other low-frequency roAp stars,
discovered from their low-amplitude radial velocity variations, include
HD\,116114 (21\,min) and $\beta$\,CrB (16.2\,min)
\citep{2005MNRAS.358..665E,2007MNRAS.380..741K}. All four of these stars occupy
a similar location in the HR diagram, around $T_{\rm eff}$ = 8000~K and $\log g$ =
4.0, with KIC\,4768731 having the shortest rotation period.

\begin{figure}
\includegraphics[width=\columnwidth]{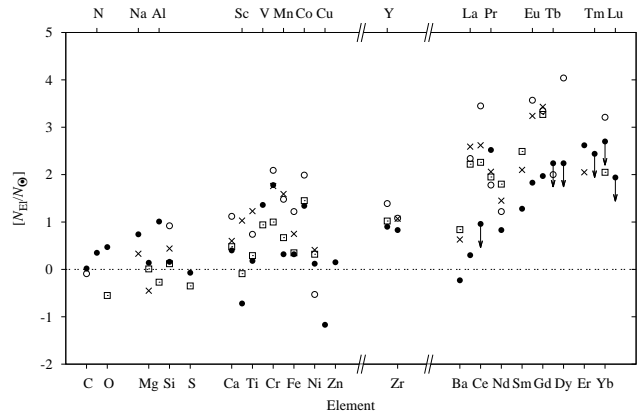}
\caption{Abundance patterns for KIC\,4768731 (filled circles), HD\,177765 (open
circles), HD\,116114 (open squares) and $\beta$\,CrB (crosses). Upper limits are indicated by downward
pointing arrows. Abundances are given relative to the Solar values from
\citet{2009ARA&A..47..481A}.}
\label{fig_abunds}
\end{figure}

Spectral lines from rare earth elements are usually strong in Ap stars
\citep{1983aspp.book.....W}. However, we find a distinct lack of significant
overabundances in rare earth elements compared to those found in other roAp
stars of similar effective temperatures: HD\,177765 \citep{2012MNRAS.421L..82A},
HD\,116114 and $\beta$\,CrB \citep{2004A&A...423..705R} (Fig.~\ref{fig_abunds}).
A similar conclusion was found for another, cooler, {\it Kepler} roAp star,
KIC\,10195926 \citep{2014MNRAS.444.1344E}.

The Mg, Si, Cr, Co, Sr, and Eu lines in KIC\,4768631 show line strength variations in
phase with rotation -- they are strongest when the star is the brightest. The Y
and Ba lines, on the other hand, vary in anti-phase with rotation --
they are strongest when the star is the faintest. Other elements, including
Ca and Fe, show little variation with rotational phase or the results are
inconclusive. The pulsation analysis has shown that the pulsation axis is in the
same plane as the abundance spots that lead to rotational light maximum, but not
completely aligned. The pulsations are coincident with light maximum, hence with
Sr\,{\sc ii} maximum, and out of phase with Y\,{\sc ii} maximum. Where the
pulsation axis, magnetic poles and spots are located is problematic, but they
appear to lie in a single plane, which is consistent with the improved oblique
pulsator model of \cite{2011A&A...536A..73B}. KIC\,4768631, however, is unusual
in having pulsation maximum at maximum rotational brightness. Many roAp stars
have pulsation maximum at rotational light minimum, which can be associated with
strong rare earth element spots.

It tends to be the slow rotators, with periods longer than one month, that have
aligned rotation and magnetic axes \citep{2000A&A...359..213L}, or the more
evolved stars, since \cite{1985A&A...148..165N} found essentially random angles
of alignment for near-ZAMS stars \citep[for a review
see][]{2014PhDT.......131M}. The former is ruled out by the 5.2\,d rotation
period, but the latter case is a possibility since the non-adiabatic modelling
suggests that the roAp-type pulsations would be expected for the star nearer the
terminal-age main-sequence (TAMS) than the ZAMS. The luminosity, and hence age,
of KIC\,4768731 is, however, uncertain due to the lack of an independent
distance determination. Furthermore, Ap\,(SrCrEu) stars are supposed to
take about half their main-sequence lifetime to develop their peculiarities
\citep{2009AJ....138...28A}, so the age is of interest. If Abt is right, then
perhaps this star is not old enough to have accumulated very anomalous rare
earth elements. Hence, KIC\,4768731 may prove to be an interesting test case
for the development of chemical peculiarities in Ap stars.

\section*{Acknowledgements}

This work made use of {\sc pyke} \citep{2012ascl.soft08004S}, a software package for the
reduction and analysis of Kepler data. This open source software project is
developed and distributed by the NASA Kepler Guest Observer Office.

Part of this work is based on observations obtained with the HERMES
spectrograph, which is supported by the Fund for Scientific Research of Flanders
(FWO), Belgium, the Research Council of K.U.Leuven, Belgium, the Fonds National
de la Recherche Scientifique (F.R.S.-FNRS), Belgium, the Royal Observatory of
Belgium, the Observatoire de Gen\'{e}e, Switzerland and the Th\"{u}ringer
Landessternwarte Tautenburg, Germany.

Calculations have been carried out in Wroc{\l}aw Centre for Networking and
Supercomputing (\url{http://www.wcss.pl}), grant No.\,214.

EN acknowledges support from the NCN grant No. 2014/13/B/ST9/00902.
SJM acknowledges research support by the Australian Research Council. Funding
for the Stellar Astrophysics Centre is provided by the Danish National Research
Foundation (grant agreement no.: DNRF106). The research is supported by the
ASTERISK project (ASTERoseismic Investigations with SONG and Kepler) funded by
the European Research Council (grant agreement no.: 267864).
MSC is supported by FCT through research grant UID/FIS/04434/2013 and through
the Investigador FCT contract of reference IF/00894/2012 and POPH/FSE (EC) by
FEDER funding through the program COMPETE. Funds for this work were provided
also by the EC, under FP7, through the project FP7-SPACE-2012-31284.
LAB wishes to thank the National Research Foundation of
South Africa for financial support.
MB is F.R.S.-FNRS Postdoctoral Researcher, Belgium
AOT acknowledges support from Sonderforschungsbereich SFB 881 ``The Milky
Way System'' (subprojects A4 and A5) of the German Research Foundation
(DFG).

\end{document}